\documentclass[12pt, epsfig, a4paper]{article}
\usepackage[dvips]{graphicx}
\usepackage{wrapfig}
\author{Yu.L.Bolotin, V.A.Cherkaskiy, G.I.Ivashkevych.}
\title{Over-barrier decay of the mixed state.}
\begin{document}
\maketitle
\begin{center}
{\footnotesize National Scientific Center "Kharkov
Institute of Physics and
Technology"\\
Akademicheskaya str., 1, 61108, Kharkov, Ukraine\\ E-mail:
giv@kipt.kharkov.ua}
\end{center}

{\footnotesize Classical escape in $2D$ Hamiltonian systems with the mixed state has been studied numerically
and analytically. The wide class of potentials with the mixed state is
presented by polinomial potentials. In
potentials, where the mixed state could be realized, i.e. the phase space contains regions of both regular and chaotic motion, escape problem has a number of new features. In particular, some
local minima become a trap with number of particles depending on
energy and other values that characterize the ensemble of
particles. Choosing the form of initial ensemble one chooses the
set of parameters that determine the number of trapped particles.}

\section{Introduction}
Escape of the trajectories from the localized regions of phase or
configuration space is an important topic in dynamics and describes
the decay of metastable states in many areas of physics, as for
instance chemical and nuclear reactions, atomic ionization and
others.
\
The problem has a rich history and a number of realizations in
different systems. Almost a century ago Sabine \cite{Sabine1} had
considered the decay of sound in concert halls and later Legrand and
Sornette \cite{Legrand} had shown that problem is equivalent to the
escape one. Corresponding decay rate is $\int {\alpha (s)ds}$ and
$\alpha (s)$ is the absorption coefficient at coordinate $s$ of
billiard boundary, $\alpha (s) = 1$ at the opening of width $\Delta$
and $\alpha (s) = 0$ elsewhere.

Another application of the escape problem links to the
nondestructive monitoring of the system \cite{Bunimovich1}. If some
system is connected to the surroundings only via small opening in
its boundary, it became possible to understand the dynamics of the
system by exploring the escaping particles. So the natural question
arises: how does escape law depend on the character of the motion?
For strongly chaotic systems exponential decay is expected
\cite{Bauer,Alt,Kokshenev}. Bauer and Bertsch \cite{Bauer}
considered the escape of the particles through the small opening in
the billiards boundary. When exploring the regular billiard, i.e.
rectangular without the scattering center, power law emerges at long
time. Qualitative understanding of the mechanism of power tails
generation is given in \cite{Bunimovich2}.

For rectangular billiard with circular scattering center inside the
decay of the initial ensemble of $N(0)$ particles is exponential and
by simple considerations one could obtain the corresponding decay
rate:

\begin{equation}\label{}
N(t) = N(0)\exp ( - \alpha t),~\alpha  = \frac{{p\Delta }}{{\pi A_c
}},
\end{equation}
here $p$ is the particles momentum, $\Delta$ - the width of the
opening and $A_c$ - the area of the billiard. As it would be shown
further, exponential decay is a common feature of the purely chaotic
systems.

\section{The mixed state}

Passing from billiards to the potential systems broadens the number
of possible situations. One-well potential is the simplest case for
considering the escape. Zhao and Du \cite{Du} have explored the
escape from Henon - Heiles potential:
\begin{equation}
U_{HH} (x,y) = \frac{{x^2  + y^2 }}{2} + xy^2  - \frac{{x^3 }}{3}
\end{equation}

At the energies $ E > 1/6$  trajectory could leave the potential
well through one of the three openings, placed symmetrically.
Numerical simulation, performed by Zhao and Du, has shown that
escape follows the exponential law. At over-saddle energies phase
space of the Henon - Heiles Hamiltonian is almost homogeneous and
motion is chaotic. Using this, the escape rate could be derived and
it fits numerical results with high accuracy. This situation is
similar to the escape from chaotic billiards. Another study of the
escape from Henon - Heiles potential was performed by
\cite{Aguirre}.

In contrast to billiards and Henon - Heiles potential, some
potentials have highly inhomogeneous phase space, that consists of
macroscopically significant components of both regular and chaotic
motions. One wide class of systems with inhomogeneous phase space is
represented by multi-well potentials. We will focus our research on
the escape in such potentials. The preliminary results are presented
in \cite{Bolotin2}. Regularity-chaos transition in multi-well
potentials has a distinctive feature, which consists in the
difference of critical energies in different local minima. This
leads to the different (either regular or chaotic) regimes of motion
in different local minima at the same energy, i.e. the ratio of
chaotic trajectories in some local minimum significantly differs
from the ratio in other minima. Such kind of the dynamics is called
the mixed state \cite{Bolotin1}. It is important to mention, that
critical energies lie below the saddle energy, i.e. the energy above
which local minima are no more separated.

\begin{figure}[h]
\hspace*{1pt}
  \begin{center}\includegraphics[scale=0.5]{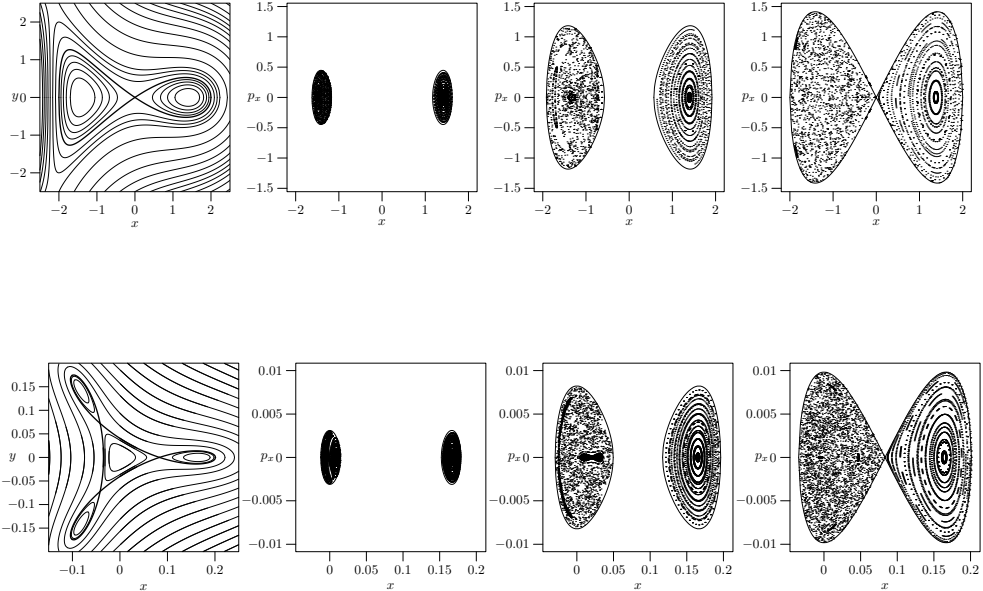}\end{center}
  \caption{\footnotesize Level lines and Poincare sections for $D_5$ (upper row) and $QO$ potentials at different energies} \label{ps_level}
\end{figure}

We will demonstrate the mixed state in two representative examples
of $2D$ multi-well potentials: the lower umbilic catastrophe $D_5$,
\begin{equation}U_{D_5 }  = 2ay^2  - x^2  + xy^2  + \frac{1}{4}x^4
\end{equation}
for $a=1.1$ and the quadrupole nuclear oscillations potential
($QO$),
\begin{equation} U_{QO} (x,y,W) = \frac{(x^2  + y^2 )}{2W} + x y^2 - \frac{1}{3}x^3  + (x^2  + y^2 )^2
\end{equation}
for $W=18$. $D_5$ potential has two local minima and three saddles
and it is the simplest potential, where where mixed state is observed.
Fig.~\ref{ps_level} shows the Poincare sections for different
energies in the considered potentials. It demonstrates the evolution of
dynamics in different local minima. At low energies motion has well
marked quasiperiodic character in both minima. As energy grows,
gradual regularity-chaos transition is observed. However,
\begin{figure}[h]
\hspace*{1pt}
  \begin{center}\includegraphics[scale=0.5]{mixed_state_all.1}\end{center}
\end{figure}
changes in features of trajectories, localized in different minima,
are sharply distinct. In the left minimum significant fraction of
trajectories becomes chaotic already at about a half of the saddle
energy, and at near saddle energy almost all initial conditions
result in chaotic trajectories. In the right minimum at the same energy
the motion remains regular and this situation is preserved up to the saddle
energy (we will call this minimum the regular one for simplicity).
Moreover, at energies above the saddle one the phase space is still
divided on chaotic and regular components, but they are not
separated in configuration space.

Earlier we have shown that the mixed state opens new possibilities
for investigations of quantum manifestation of classical
stochasticity \cite{Berezovoj}. Aim of the present work is to study
the classical escape from separated local minima, realizing the
mixed state. We show that escape from such local minima has all
above mentioned properties of the decay of chaotic systems and also
a diversity of principally new features, representing an interesting
topic for conceptual understanding of chaotic dynamics and for
applications as well. We are interested in both the "first passage"
effects and dynamical equilibrium setup for the finite motion (for
example, in $QO$ potential). It is important to stress, that though
we study the process of escape from concrete local minimum, the
over-barrier case of the mixed state has specific memory: general
phase space structure at supersaddle energies is determined by the
characteristics of the motion in all other local minima.

\section{Decay of the uniformly distributed ensemble}

\begin{figure}
\hspace*{1pt}
  \begin{center}\includegraphics[scale=0.8]{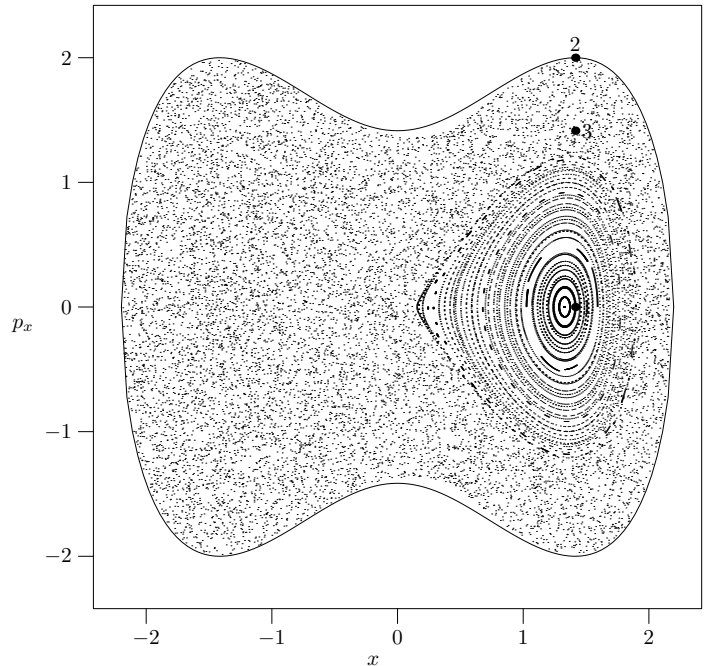}\end{center}
  \caption{\footnotesize Poincare section for $D_5$ potential at $E=1.0$. Initial conditions for quantum computations are presented} \label{ps_full}
\end{figure}

At the energies above the saddle, i.e. $E>E_S$, different components
are not separated in the configuration space. Fig.~\ref{ps_full}
represents the Poincare section for $D_5$ potential at supersaddle
energy. The ''chaotic sea'' stretches on whole accessible area,
while regular island in the right well is localized. This means that,
been initially localized in right well, chaotic trajectories could
leave the well and regular ones remain trapped, i.e. the decay of the mixed
state occurs. Therefore we will explore the escape of the particles
from the right well of $D_5$ and peripheral wells of $QO$ potential.

Numerical simulation of the escape process in these potentials implies
three steps. At the first stage we select initial distribution of
the particles inside the well. Then direct numerical integration of
the equations of motion for all particles is performed and we
extract $N(t)/N(0)$---relative number of particles in the well.
Using this function one can calculate escape rate and part of the
trapped trajectories.
One remark should be made about the first step of the numerical
simulation. Initial distribution, in general, determines the ratio
between regular and chaotic trajectories and hence it should be
physically motivated. One chance is to distribute particles
uniformly in all classically allowed configuration space and another---to put all particles at the same point. Second case emulates
injection of particles to the well. In both cases momentum is
calculated using the energy conservation (in the second it will be
the same for all particles) and its direction is uniformly
distributed in $\left[ {0,2\pi } \right]$. These initial
distributions present quite simple extreme cases of real
distributions. Uniform distribution will be illustrated with
$U_{D_5}$ and "point" distribution - with $U_{QO}$.

Phase space density for uniform initial distribution is
\begin{equation}\label{ro_eq} \rho (E) = \frac{1}{{2\pi S(E)}}
\end{equation}
where $S(E)$ is the area of classically allowed space:
\begin{equation}
S(E) = \int\limits_{x > x_S } {dxdy\Theta (E - U(x,y))}
\end{equation}

Numerical simulation reveals some substantial features of escape
process for this initial distribution. Fig.~\ref{N_t_D5}
demonstrates the normalized number of particles in the well as a
function of time. Decay law has three important features:
\begin{figure}
\hspace*{1pt}
  \begin{center}\includegraphics[scale=0.8]{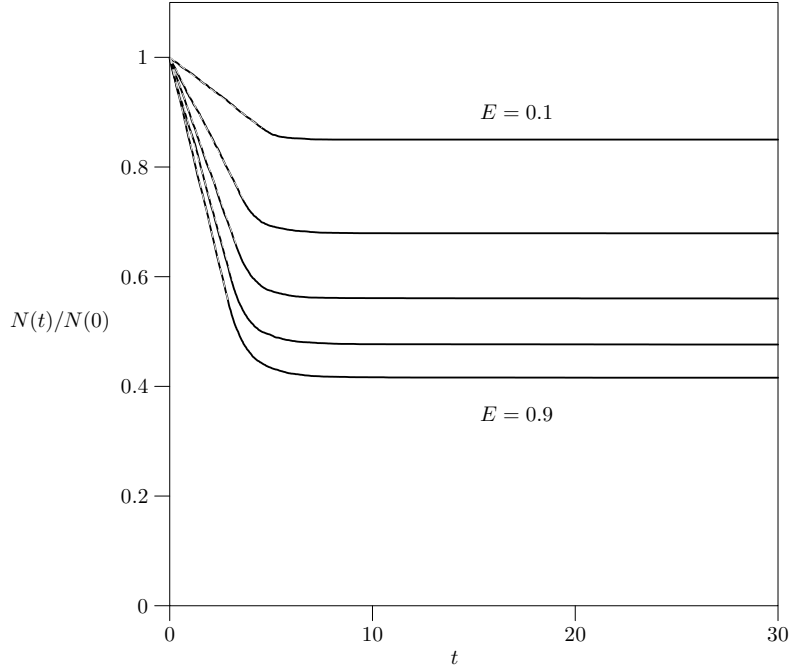}\end{center}
  \caption{\footnotesize Decay law $N(t)/N(0)$ for $D_5$ potential at different energies. $E_i=0.1+0.2k,~k=0..4$} \label{N_t_D5}
\end{figure}
\begin{itemize}
    \item saturation at $t\rightarrow\infty$:

    \begin{equation}
     N(t \to \infty ) = \rho _\infty  N_0
    \end{equation}

    Because of uniform initial distribution, quantity $\rho _\infty$ is a relative phase volume, occupied by trapped
    trajectories.
    \item initial linear decrease - from $0$ to some $\tau
(E)$:

    \begin{equation}\label{linear_law_eq}
    N(t)/N_0  = 1 - \alpha ^{(l)} t
    \end{equation}

    \item exponential decrease at $t > \tau (E)$:

     \begin{equation}\label{exp_law_eq}
      N(t)/N_0  = \rho _\infty   + C\exp ( - \alpha ^{(e)}t)
      \end{equation}

\end{itemize}
\begin{figure}
\hspace*{1pt}
  \begin{center}\includegraphics[scale=0.7]{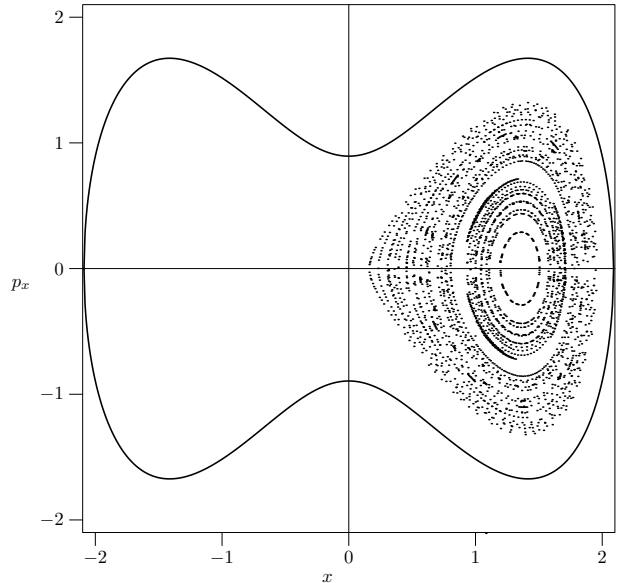}\end{center}
  \caption{\footnotesize Poincare section for trapped trajectories. These trajectories form regular island, i.e. they are regular} \label{plato}
\end{figure}
\begin{figure}
\hspace*{1pt}
  \begin{center}\includegraphics[scale=0.7]{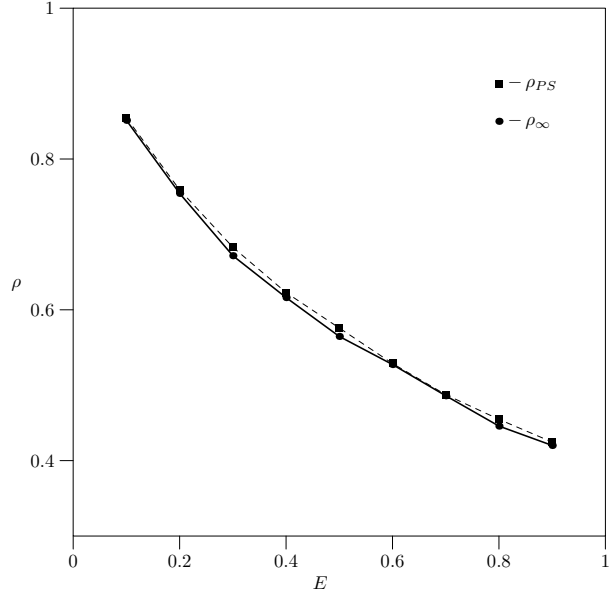}\end{center}
  \caption{\footnotesize $\rho$, the relative area of the regular island in Poincare section, is shown by squares and $\rho _\infty  (E)$ is shown by circles} \label{ro_D5}
\end{figure}

Fig.~\ref{plato} presents the Poincare section for trapped
trajectories. Obviously, regular island in the right well is formed by
trapped, i.e regular, trajectories. Function $\rho _\infty  (E)$
is demonstrated on Fig.~\ref{ro_D5}. This is decreasing function,
as expected from physical considerations---number of regular
trajectories decreases as energy increases.

It is interesting to mention, that there exists correlation between
$\rho _\infty (E)$ and relative area of regular island in Poincare
section $\rho$. While relative part of trapped trajectories is
linked to the volume of four-dimensional phase space, occupied by
regular trajectories, Poincare section is two-dimensional and thus
there is no argument to expect the precise coincidence of $\rho$ and
$\rho _\infty (E)$.

To calculate the relative area of regular island in Poincare section
we first determine the border of the island through numerical
integration of equation of motion and calculate the area inside it.
Then this area was divided by total area, determined by the conditions
$x>0,~p^2>0$. In spite of topological inequivalence, $\rho$ and
$\rho _\infty (E)$ are very close to each other. This means that one
could determine and control the part of trapped trajectories using
only the Poincare section.

Linear part of the decay law is more pronounced in comparison with
pure ensemble \cite{Du}. In the time interval between 0 and $\tau
(E)$ decay has the form (\ref{linear_law_eq}). Thus, there are two
quantities describing linear decay---its duration $\tau (E)$ and
corresponding escape rate $\alpha ^{(l)}$.

After linear part, at $t > \tau (E)$, decay law has exponential form
(\ref{exp_law_eq}). It's important that (\ref{linear_law_eq}) is
not a linear approximation of (\ref{exp_law_eq}), and linear decay
has independent nature. This means that $\rho _\infty + C \ne 1$.
Moreover, decay law of the form (\ref{linear_law_eq}) precisely works
up to time $t\sim\tau (E)$ and in this interval it differs
substantially from corresponding exponential law $\rho _\infty + (1
- \rho _\infty )\exp ( - \alpha ^{(l)} t)$.

Let's now calculate $\tau (E)$. From analysis of numerical
calculation one can derive that this time corresponds to the time
of one-dimensional, along $y=0$, motion from the saddle to
opposite side of the well and back. Thus $\tau (E)$ has the form:
\begin{equation}
\tau _{D_5 } (E) = 2\int\limits_0^{\sqrt {2(1 + \sqrt E )} }
{\frac{{dx}}{{\left| v \right|}}}  = \frac{{\sqrt 2
}}{{E^{\frac{{1}}{{4}}} }}K\left(\sqrt {\frac{{1 + \frac{1}{{\sqrt E
}}}}{2}}\right)
\end{equation}
\begin{equation}
\tau _{QO} (E) = 12\left(\frac{{E_S }}{E}\right)^{\frac{1}{4}} K\left(\sqrt
{\frac{{1 + \sqrt {\frac{{E_S }}{E}} }}{2}}\right) = 6\sqrt 2 \tau
_{D_5 } \left(\frac{{E_S }}{E}\right)
\end{equation}
where $K(k)$ is full elliptic integral of the first type and $E_s =
1/12^4$---saddle energy in $QO$ potential at $W=18$.

Such nature of $\tau (E)$ and analysis of linearly escaping
trajectories allow to conclude that linear decay corresponds to
the escape of trajectories that move along $y=0$ and cross the
well not more than two times. Correspondingly, trajectories which
initially move toward saddle along $y=0$ escape first, and then
escape occur for trajectories moving to the opposite part of the
well.

$\alpha ^{(l)}$ could be calculated via averaging of the flow
through the saddle:
\begin{equation}\label{alpha_gen}
\alpha ^{(l)} (E) = \rho (E) \int\limits_{x = x_S } {dy} \int_{ -
{\pi  \mathord{\left/
 {\vphantom {\pi  2}} \right.
 \kern-\nulldelimiterspace} 2}}^{{\pi  \mathord{\left/
 {\vphantom {\pi  2}} \right.
 \kern-\nulldelimiterspace} 2}} {d\theta \left| v \right|\cos \theta
 } \label{linear_rate}
\end{equation}
This procedure is the same as in \cite{Du}. Using the density
(\ref{ro_eq}) and integrating the (\ref{alpha_gen}) we obtain the
expression for linear escape rate:
\begin{equation}
\alpha _{D_5 }^{(l)} (E) = \frac{{E}}{{2\sqrt a S_{D_5 } (E)}}
\end{equation}
\begin{equation}
 \alpha _{QO}^{(l)} (E) = \frac{{\sqrt[4]{\varepsilon }}}{{12\pi S_{QO} (E)}}\\\times \left\{ {(16\sqrt \varepsilon   + 1)K\left(\sqrt {\frac{{1 - \frac{1}{{16\sqrt \varepsilon  }}}}{2}}\right) - 2E\left(\sqrt {\frac{{1 - \frac{1}{{16\sqrt \varepsilon  }}}}{2}}\right)} \right\} \\
\end{equation}
Exponential decrease of $N(t)/N(0)$, as it could be understood
from the analysis of escaped trajectories, corresponds to the
leaving of sticking orbits, i.e. those chaotic trajectories which
moved in the vicinity of the regular island in the Poincare section.

Energy is the parameter which determines the part of trapped
trajectories for the uniform ensemble. Changing the energy of
ensemble one could trap the given number of particles.

\begin{figure}[h]
  \begin{center}\includegraphics[scale=0.5]{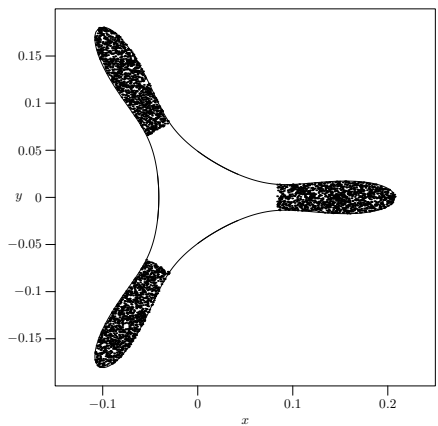}\end{center}
  \caption{\footnotesize Initial ensemble for extraction of asymptotic distribution.} \label{init_distr}
\end{figure}

\begin{figure}[h]
  \begin{center}\includegraphics[scale=0.5]{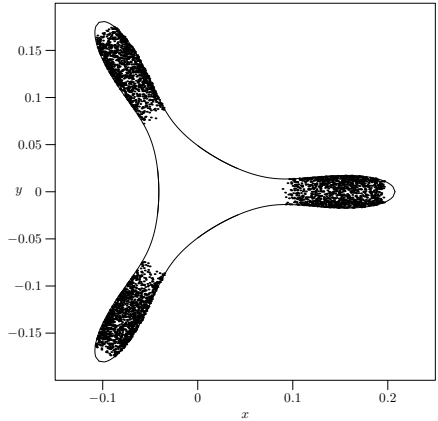}\end{center}
  \caption{\footnotesize Asymptotic distribution of trapped particles.} \label{reg_distr}
\end{figure}

\begin{figure}[h]
  \begin{center}\includegraphics[scale=0.5]{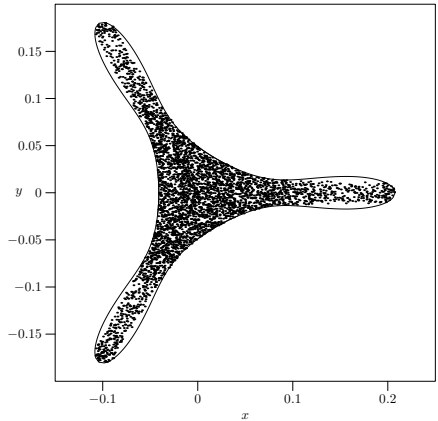}\end{center}
  \caption{\footnotesize Asymptotic distribution of free particles.} \label{ch_distr}
\end{figure}

To illustrate the splitting of initial ensemble on regular and
chaotic components one needs to plot the ensemble in $(x,y)$ plane at
time $t\gg\tau(E)$. It is convenient to plot the asymptotic
evolution of the initial ensemble in $QO$ potential because of finite
character of motion in it. At $t=0$ particles are equally
distributed between peripheral minima. Fig.~\ref{init_distr}
represents this initial ensemble.

The ensemble splits during the evolution in time---regular trajectories
remain trapped in peripheral minima, while chaotic trajectories
cover almost entire accessible configuration space. Integrating
equations of motion for all trajectories in the initial ensemble for
enough long time (in fact---for time much greater than the typical
escape time) we obtain the particles positions corresponding to
asymptotic distribution. We have calculated the asymptotic
distribution for energy $E=1.5 E_S$ ($E_S=1/12^4$ -
saddle energy in $QO$ potential), while integration was performed
for $t=150$ (for this energy $\tau(E)=28.395$). Fig.~\ref{reg_distr}
shows the asymptotic distribution of trapped particles. At enough
large time these particles tend to accumulate closer to the center
of the well.

The corresponding distribution of free particles is represented at
fig.~\ref{ch_distr}. As it was mentioned above, free particles cover
entire central minimum and, according to their character, the area
near the plane $y=0$ in peripheral minima. If free particles are
removed in some way after their leave the peripheral minima (this
is, of course, correct for right well of $D_5$ and any well of same
topology), we obtain pure regular ensemble inside the well, i.e. the 
initial mixed ensemble splits.

\section{Point ensemble}
For simplicity we will consider the point distribution with $y=0$.
Thus the governing parameter is $x_0$---$x$--coordinate of the point of
injection. Corresponding density has the form:
\begin{equation}
\label{ro_in}
\rho \left( {x,y,p,\varphi } \right) = \frac{{\delta
\left\{ {x - x_0 } \right\}\delta \left\{ y \right\}\delta \left\{
{p - \sqrt {2(E - U(x_0 ,0))} } \right\}}}{{2\pi \sqrt {2(E -
U(x_0 ,0))} }}
\end{equation}

Numerical procedure is identical to that for uniform ensemble, but
for distribution ~(\ref{ro_in}) one could made general conclusion
about the character of decay even without numerical integration.

For this one needs to consider the representation of initial
distribution on the Poincare section:
\begin{equation}
\rho _{PS} (x,p_x ) = \frac{{\delta \left\{ {x - x_0 }
\right\}\Theta \left\{ {p_x  - \sqrt {2(E - U(x_0 ,0))} }
\right\}\Theta \left\{ {\sqrt {2(E - U(x_0 ,0))}  + p_x }
\right\}}}{{2\sqrt {2(E - U(x_0 ,0))} }}
\end{equation}
where $\Theta \left\{ a \right\}$ is a step function. Lets denote:
\begin{equation}
p_x^{(\max)}  = \sqrt {2(E - U(x_0,0))}.
\end{equation}
Thus in the Poincare section initial ensemble occupies the interval $x
= x_0 ,p_x \in \left[ { - p_x^{(\max )} ,p_x^{(\max )} } \right]$.
Edges of this interval correspond to the momentum directions
$\varphi = \pi ,0$. In the over--barrier case this interval crosses the
regular island in the points $(x_0,p_x^{(reg)})$ and
$(x_0,-p_x^{(reg)})$. Obviously, particles with $p_x$ in the interval
$\left[ { - p_x^{(reg)} ,p_x^{(reg)} } \right]$ could not leave
the well. Value
\begin{equation}
\varphi _{\max } (E,x_0 ) = 2\arccos \left(\frac{{p_x^{(reg)}
}}{{p_x^{(\max )} }}\right)
\end{equation}
defines the cone of directions which could leave the well. In other
words, particles with $p_y$ which is greater than some maximum
value are trapped. Thus first conclusion about decay of point
ensemble implies the existence of escape cone and this feature
reveals the role of transversal momenta in the escape process.
\begin{figure}
\hspace*{1pt}
  \begin{center}\includegraphics[scale=0.7]{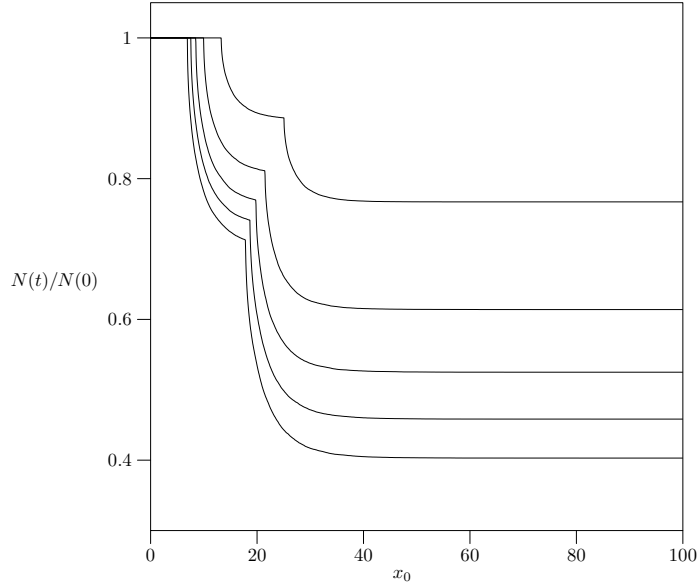}\end{center}
  \caption{\footnotesize Normalized particle number in the well for point ensemble at different energy and $x_0=0.16$} \label{nqo}
\end{figure}

The second feature consists in the fact that decay begins at the
time $\tau_1$, which corresponds to the time of motion of particle
with momentum $p_x=-p$ from point $x_0$ to the saddle:
\begin{equation}
\label{tau1} \tau _1  = \int\limits_{x_S}^{x_0} {\frac{{dx}}{{p_x
}}} = \int\limits_{x_S}^{x_0} {\frac{{dx}}{{\sqrt {2(E - U(x,0))}
}}}
\end{equation}
Moreover, escape is a two-stage process due to existence of escape
cone. At the second stage the particles, moving toward the well boundary
in the escape cone, leave.

Numerical procedure implies the determination of the $\varphi
_{\max } (x_0 ,E)$, $\tau _1 (x_0 ,E)$, and $N(t=\infty)$, i.e.
the number of particles in the well. Due to uniform distribution
of momenta directions one has the relation:
\begin{equation}
\rho _\infty  (x_0 ,E) \equiv \frac{{N(\infty )}}{{N(0)}} = 1 -
\frac{{\varphi _{\max } (x_0 ,E)}}{\pi }
\end{equation}

We will illustrate the above consideration for $QO$ potential.
Energy is normalized to the saddle energy, $E_S=1/144^2$:
$E=\varepsilon E_S$. Initial ensemble is localized in peripheral
minimum with $x_{min}=1/6$ and corresponding saddle $x_S=1/12$.
Fig.~\ref{nqo} represents the normalized number of particles in
the well as a function of time for different values of energy and
$x_0=0.16$.

\begin{figure}
\hspace*{1pt}
  \begin{center}\includegraphics[scale=0.7]{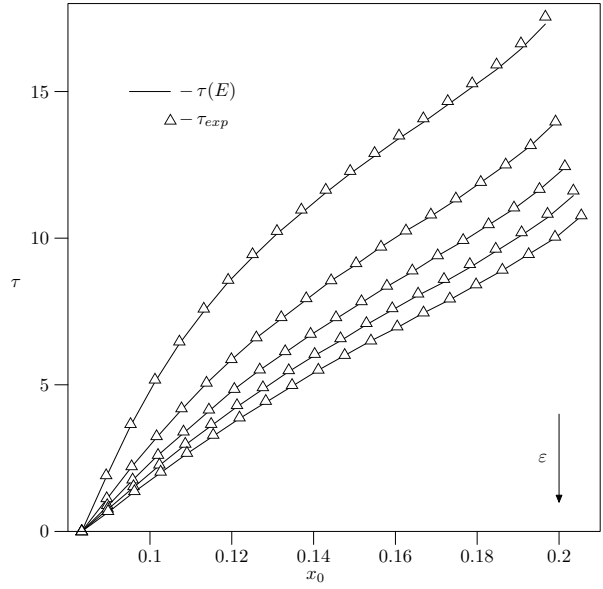}\end{center}
  \caption{\footnotesize Numerically obtained $\tau_1$ and analytically calculated} \label{tau_fig}
\end{figure}

Numerically obtained $\tau_1$ could be compared to analytical
value:
\begin{equation}
\tau _1  = \int\limits_{1/12}^{x_0} {\frac{{dx}}{{p_x }}}  =
\int\limits_{1/12}^{x_0} {\frac{{dx}}{{\sqrt {2(\varepsilon E_S -
U_{QO} (x,0))} }}}
\end{equation}
Numerical and analytical value of first escape time are presented
at fig.~\ref{tau_fig}

\begin{figure}
\hspace*{1pt}
  \begin{center}\includegraphics[scale=0.7]{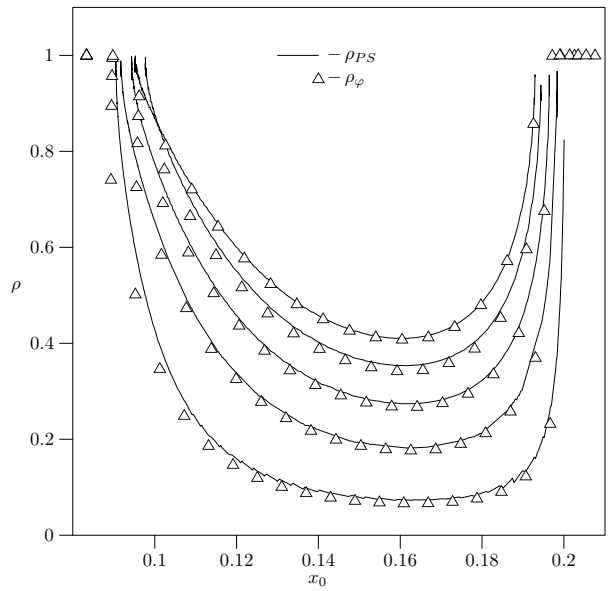}\end{center}
  \caption{\footnotesize $\rho_{PS}$ and $\rho_\varphi$ for different energies and injection points} \label{ps_fi_fig}
\end{figure}

The most interesting question is the correspondence between escape
cone angle, number of trapped particles and linear part of regular
island in Poincare section. Fig.\ref{ps_fi_fig} represents the
quantities
\begin{equation}
\label{ro_ps} \rho _{PS}  = 1 - \frac{{p_x^{(reg)} (x,\varepsilon
)}}{{p_x^{(\max )} (x,\varepsilon )}}
\end{equation}
and
\begin{equation}
\label{ro_fi} \rho _\varphi   = 1 - \cos
({\raise0.7ex\hbox{${\varphi _{\max } (x,\varepsilon )}$}
\!\mathord{\left/
 {\vphantom {{\varphi _{\max } (x,\varepsilon )} 2}}\right.\kern-\nulldelimiterspace}
\!\lower0.7ex\hbox{$2$}})
\end{equation}
First is a linear part of regular island in the section and second is the corresponding expression through the escape cone angle.

This angle could be determined during numerical simulation. Thus,
Poincare section could be used to determine the angle of the escape
cone. On the other hand, this angle
\begin{figure}
\hspace*{1pt}
  \begin{center}\includegraphics[scale=0.7]{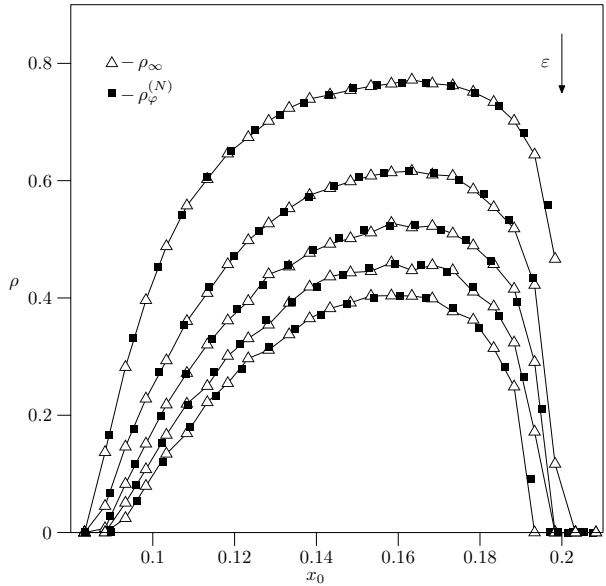}\end{center}
  \caption{\footnotesize $\rho_{N}$ and $\rho_\varphi$ for different energies and injection points} \label{n_fi_fig}
\end{figure}is connected with the part of trapped trajectories. To demonstrate
this connection one needs to compare
\begin{equation}
\label{ro_fi_n}\rho _\varphi ^{(N)}  = 1 - \frac{{\varphi _{\max }
(x,\varepsilon )}}{\pi }
\end{equation}
with $\rho_\infty$. Corresponding data are presented at
fig.~\ref{n_fi_fig}. This analysis allows to conclude that point
ensemble differs substantially from the uniform when considering
the escape.

In uniform ensemble Poincare section gives only estimate (nevertheless very
accurate) of trapped particles number, while in the point ensemble one
could calculate not only part of trapped particles, but also the
escape cone angle, using only Poincare section. Point ensemble
allows dual control of trapped particles part using the energy and
injection point. One even needs not to compute the entire Poincare
section, but only the boundary of regular island,
$p_x^{(reg)}(x_0,\varepsilon)$, and then use the relation:
\begin{equation}
\rho_\infty  = 1 - \frac{2\arccos \left(\frac{{p_x^{(reg)}
}}{{p_x^{(\max)} }}\right)}{\pi}
\end{equation}
The procedure of calculating the part of trapped particles, thus, implies three steps. First of all one needs to calculate the boundary of the regular island in the Poincare section. After that, the $p_x^{(reg)}$ and $p_x^{(\max)}$ should be determined. Now value $\rho_{\infty}$ could be calculated. 

\section{Quantum escape problem}
Now we consider the over-barrier decay of the mixed state from the
quantum-mechanical point of view. In semiclassical limit, temporal
evolution of an initial state in form of the minimum uncertainty
Gaussian wave packet
\[\Psi_G(x,y;x_0,y_0,p_{x0},p_{y0})=\frac{1}{\sqrt{\pi\sigma_x\sigma_y}}
e^{-\frac{(x-x_0)^2}{2\sigma_x^2}-\frac{(y-y_0)^2}{2\sigma_y^2}
+\frac{ip_{x0}}{\hbar}(x-\frac{x_0}{2})+\frac{ip_{y0}}{\hbar}(y-\frac{y_0}{2})}\]
represents the quantum analogue for classical motion of a point
particle with initial condition $(x_0,y_0,p_{x0},p_{y0})$. The
quantum-classical correspondence between the wave packet motion and
the classical trajectory preserves quite a long time until the wave
packet spreads.

We made numerical simulations for time evolution of the Gaussian
wave packets for three different initial conditions in the $D_5$
potential (3) with $a=1.1$. The following wave packets parameters
\[\hbar=0.01,\ \sigma_x=\sqrt{\frac \hbar 2}\simeq0.07,\
\sigma_y=\sqrt{\frac{\hbar}{2\sqrt{a+\frac{1}{\sqrt2}}}}\simeq0.06,\
x_0=\sqrt2,\ y_0=0\] and $p_0=\sqrt{p_{x0}^2+p_{y0}^2}=2$ were the
same for all the three initial states, so all of them started from
the right local minimum of the potential with initial energy $E=1$,
which is twice higher than the barrier height. (Recall that in the
$D_5$ potential (3) $E_{min}=-1$ and $E_S=0$). The only difference
was in the direction of initial momentum: we considered the cases
\[\varphi_0=0,\ \frac\pi4,\ \frac\pi2.\]

The initial condition with $\varphi_0=\pi/2$ falls near the center
of the principal stability island and therefore it corresponds to
regular classical trajectory trapped in the right potential well.
Quantum autocorrelation function
\begin{equation}\label{ac}P(t)=\int
dxdy\Psi(t=0)\Psi(t)\end{equation} manifests quasiperiodic nature of
the corresponding classical trajectory: the sharp peaks indicate
periodically repeated recurrences of the wave packet to the initial
state, and rather high amplitude of the peaks shows that there is
almost no spreading of the wave packet. The reason of that slow
spreading is the fact that even for highly over-barrier energies the
regular classical motion near the potential minimum is still pretty
close to that in two-dimensional harmonic potential, as it is seen
for example in the characteristic structure of the stability island
on the Poincar\'e section. For the considered case of the right
local minimum in the $D_5$ potential the corresponding quadratic
(harmonic) approximation reads
\begin{equation}U(x,y)\simeq\frac{\omega_x^2(x-\sqrt2)^2+\omega_y^2
y^2}{2}-1\label{ha}
\end{equation} with $\omega_x=2$ and
$\omega_y=2\sqrt{a+1/\sqrt2}\simeq2.7$ for $a=1.1$. Therefore the
trapped quasiperiodic trajectories actually are very close to the
Lissajoux figures --- simple superpositions of harmonic oscillations
in perpendicular directions. Remarkably, the considered gaussian
wave packet coincides with the coherent state for the harmonic
oscillator potential (\ref{ha}) --- the exact non-spreading solution
of time depending Schr\"odinger equation --- that is why it is
"almost" non-spreading near the minimum of the $D_5$ potential. As
the considered initial condition $\varphi_0=\pi/2$ leads to almost
one-dimensional motion (along the $y$-axis), only one frequency
$\omega_y\simeq2.7$ is manifested in the autocorrelation function
$P(t)$ (\ref{ac}).

The initial condition $\varphi_0=0$ also corresponds to a regular
trajectory --- the periodic one-dimensional motion along the
$x$-axis. The autocorrelation function (\ref{ac}) for the
corresponding wave packet clearly shows the same periodicity of
recurrences, but decreasing amplitude of the corresponding peaks
reveals much faster spreading than in the former case
$\varphi_0=\pi/2$. Clear explanation for it is that any harmonic
approximation is no more valid
\begin{figure}
\hspace*{1pt}
  \begin{center}\includegraphics[scale=0.6]{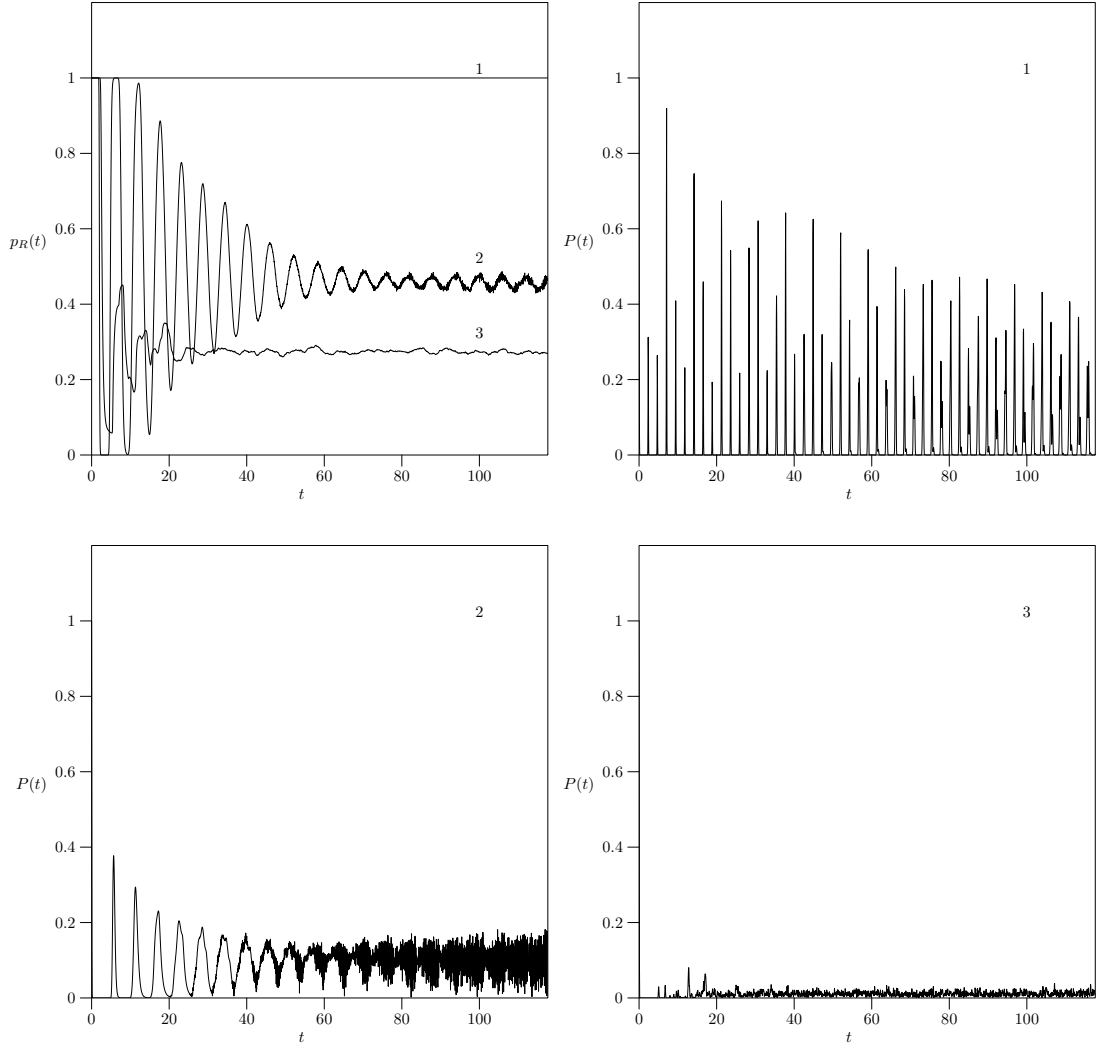}\end{center}
  \caption{\footnotesize Survival probabilities and autocorellation functions for considered initial conditions} \label{quantum}
\end{figure}
for the trajectory, and the Gaussian
wave packet is not already a good approximation for the exact time
dependent solution. It is worth noting that the wave packet for the
case $\varphi_0=0$ undergoes only one-dimensional spreading ---
along the $x$-axis, while it remains well localized in the
$y$-direction.

The case with $\varphi_0=\pi/4$ sharply differs from the two
formers, as it corresponds to a chaotic trajectory. Accordingly the
autocorrelation function shows almost absent recurrences and even
faster spreading of the wave packet. This time the wave packet
spreads already in two dimensions --- over all the chaotic sea.

Survival probability for the right potential well
\[p_R(t)=\int\limits_{x>0} dxdy |\Psi|^2\] is a natural quantum
analogue of the classical quantity $N(t)/N(0)$. The results of
numerical simulations for all the three considered cases are
presented on fig.\ref{quantum}. Naturally, $p_R\equiv1$ for the trapped state
($\varphi_0=0$), as well as for the two others for
$t<t_{cl}\simeq2.5$ --- the classical escape time which is almost
the same for both the wave packets. Periodicity of the trajectory
with $\varphi_0=0$ perfectly manifests again in the periodic
character of the $p_R(t)$ for the corresponding wave packet.
Graduate decay of the oscillations amplitude is due to the
spreading. Saturation of the survival probability at value
$p_r\simeq0.5$ confirms that the spreading is one-dimensional in
that case: the probability density $|\Psi|^2$ gets uniformly
distributed along the one-dimensional trajectory, which lies between
the points $x_{1,2}=\pm\sqrt2\sqrt{1+\sqrt2}$ (recall that $E=1$),
so exactly one half of $|\Psi|^2$ falls into each local minimum.

On the case of the chaotic motion $\varphi_0=\pi/4$ the survival
probability is aperiodic and quickly saturates at the value
$p_R\simeq0.3$. It is in accordance with our arguments of fast
spreading of the wave packet to uniform two-dimensional distribution
of the probability density $|\Psi|^2$  over the chaotic sea ---
indeed, exactly about $30\%$ of the chaotic sea amounts for the
right minimum, as it can be seen from the Poincar\'e section. If we calculate the ratio of the chaotic sea in the right minimum to the whole area of the sea in Poincar\'e section, the exact value would be $0.39$. 

Our preliminary considerations of three principal types of initial
conditions show generally good coincidence of the quantum and
classical results for the decay of the mixed state. Specifically
quantum effects in the escape problem, such as the resonance barrier
penetration and the chaos-assisted dynamical tunneling require
further analysis, which will be published elsewhere.

\section{Conclusions}

Investigation of the escape from localized areas of configuration
space in the existence of the mixed state is presented. When the mixed state
is present in the system, it is possible to "trap" given number of
particles in the well. We have considered two possible initial
distributions.

For uniform distribution escape law splits into three sections.
First section corresponds to linear decay, second - to exponential
and third forms the plato, which corresponds to trapped particles.
Number of trapped particles depends only on energy.

In the point ensemble case the escape is a two--stage process and
number of trapped particles depends not only on energy, but on
coordinate of injection point too. Only trajectories with
direction of initial momenta in some cone could escape. Angle of
escape cone is connected with linear part of regular island in
the Poincare section and number of trapped trajectories. Moreover,
only edge of regular island is necessary to compute the number of
trapped particles.

\end{document}